\documentclass[a4paper,11pt]{article}

\usepackage{jcappub}
\usepackage[normalem]{ulem}

\usepackage{slashed}
\usepackage{bm}
\usepackage{latexsym,amssymb,amsmath,float,url,mathrsfs}
\usepackage{latexsym}
\usepackage{epstopdf}
\usepackage{amsfonts}
\usepackage{amsmath}
\usepackage{amssymb}
\usepackage{comment}
\usepackage{subfigure}
\usepackage{natbib}
\usepackage{hyperref}
\usepackage{pifont}
\usepackage{blindtext}
\usepackage{adjustbox}
\usepackage{multirow}
\usepackage{tabularx}
\usepackage[table]{xcolor}
\usepackage{orcidlink}
\usepackage{tikz}
\usetikzlibrary{tikzmark}
\usepackage{amssymb}
\usepackage{bbold}
\usepackage{blkarray}
\usepackage{graphicx}
\usepackage{amsfonts}
\usepackage{soul}
\usepackage{amssymb}
\usepackage{amsmath}
\usepackage{cancel}
\usepackage{tcolorbox}
\def\draftlabel#1{{\@bsphack\if@filesw {\let\thepage\relax
			\xdef\@gtempa{\write\@auxout{\string
					\newlabel{#1}{{\@currentlabel}{\thepage}}}}}\@gtempa
		\if@nobreak \ifvmode\nobreak\fi\fi\fi\@esphack}
	\gdef\@eqnlabel{#1}}
\def\@eqnlabel{}
\def\@vacuum{}
\def\draftmarginnote#1{\marginpar{\raggedright\scriptsize\tt#1}}

\def\draft{\oddsidemargin -.5truein
		\def\@oddfoot{\sl preliminary draft \hfil
			\rm\thepage\hfil\sl\today\quad\militarytime}
		\let\@evenfoot\@oddfoot \overfullrule 3pt
		\let\label=\draftlabel
		\let\marginnote=\draftmarginnote
		\def\@eqnnum{(\theequation)\rlap{\kern\marginparsep\tt\@eqnlabel}%
			\global\let\@eqnlabel\@vacuum}  }
	
	\def\preprint{\twocolumn\sloppy\flushbottom\parindent 2em
		\leftmargini 2em\leftmarginv .5em\leftmarginvi .5em
		\oddsidemargin -.5in    \evensidemargin -.5in
		\columnsep .4in \footheight 0pt
		\textwidth 10.in        \topmargin  -.4in
		\headheight 12pt \topskip .4in
		\textheight 6.9in \footskip 0pt
		\def\@oddhead{\thepage\hfil\addtocounter{page}{1}\thepage}
		\let\@evenhead\@oddhead \def\@oddfoot{} \def\@evenfoot{} }
	
	\def\numberbysection{\@addtoreset{equation}{section}
		\def\theequation{\thesection.\arabic{equation}}}
	
	\def\underline#1{\relax\ifmmode\@@underline#1\else
		$\@@underline{\hbox{#1}}$\relax\fi}

	\def\titlepage{\@restonecolfalse\if@twocolumn\@restonecoltrue\onecolumn
		\else \newpage \fi \thispagestyle{empty}\c@page\z@
		\def\thefootnote{\fnsymbol{footnote}} }
	
	\def\endtitlepage{\if@restonecol\twocolumn \else \newpage \fi
		\def\thefootnote{\arabic{footnote}}
		\setcounter{footnote}{0}}  

\catcode`@=12
\relax

\def\figcap{\section*{Figure Captions\markboth
		{FIGURECAPTIONS}{FIGURECAPTIONS}}\list
	{Figure \arabic{enumi}:\hfill}{\settowidth\labelwidth{Figure
			999:}
		\leftmargin\labelwidth
		\advance\leftmargin\labelsep\usecounter{enumi}}}
 \relax
\def\tablecap{\section*{Table Captions\markboth
		{TABLECAPTIONS}{TABLECAPTIONS}}\list
	{Table \arabic{enumi}:\hfill}{\settowidth\labelwidth{Table
			999:}
		\leftmargin\labelwidth
		\advance\leftmargin\labelsep\usecounter{enumi}}}
 \relax
\def\reflist{\section*{References\markboth
		{REFLIST}{REFLIST}}\list
	{[\arabic{enumi}]\hfill}{\settowidth\labelwidth{[999]}
		\leftmargin\labelwidth
		\advance\leftmargin\labelsep\usecounter{enumi}}}
 \relax

	\makeatletter
	\newcounter{pubctr}
	\def\publist{\@ifnextchar[{\@publist}{\@@publist}}
	\def\@publist[#1]{\list
		{[\arabic{pubctr}]\hfill}{\settowidth\labelwidth{[999]}
			\leftmargin\labelwidth
			\advance\leftmargin\labelsep
			\@nmbrlisttrue\def\@listctr{pubctr}
			\setcounter{pubctr}{#1}\addtocounter{pubctr}{-1}}}
	\def\@@publist{\list
		{[\arabic{pubctr}]\hfill}{\settowidth\labelwidth{[999]}
			\leftmargin\labelwidth
			\advance\leftmargin\labelsep
			\@nmbrlisttrue\def\@listctr{pubctr}}}
	 \relax
	\makeatother
	\newskip\humongous \humongous=0pt plus 1000pt minus 1000pt

	\newif\ifdtup

	\relax
	
	\def\be{\begin{equation}}
		\def\ee{\end{equation}}
	\def\ba{\begin{eqnarray}}
		\def\ea{\end{eqnarray}}

	\def\cR{{\cal R}}

	  \def\cR{{\cal R}}

\newcommand{\keff}{\kappa_{\rm ind}}
\newcommand{\Teff}{T_{\rm ind}}

\newcommand{\prt}[1]{{\left( {#1} \right)}}

	\newcommand{\ff}{\frac}

	\def \black hole { {\bar h} }

	\def\IR{\relax{\rm I\kern-.18em R}}
	\def\IL{\relax{\rm I\kern-.18em L}}
	
	\def\inv{^{\raise.15ex\hbox{${\scriptscriptstyle -}$}\kern-.05em 1}}

	\def\bea{\begin{eqnarray}}
		\def\eea{\end{eqnarray}}

	\newcommand{\la}[1]{\label{#1}}

	  \def\D{\Delta}

	\def\th{\theta}

	\definecolor{markcolor2}{rgb}{1,0,0}
	
	\definecolor{markcolor3}{rgb}{0,1,0}

    \def\dd{{\rm d}}

\usepackage{graphics,appendix,afterpage,makecell} 

\def\eq#1{{Eq.~(\ref{#1})}}

\def\fig#1{{Fig.~\ref{#1}}}

\definecolor{oucrimsonred}{rgb}{0.6, 0.0, 0.0}
\definecolor{persianblue}{rgb}{0.11, 0.22, 0.73}
\definecolor{forestgreen}{rgb}{0.13,0.35,0.13}
\definecolor{lightgray}{rgb}{0.83, 0.83, 0.83}
 \hypersetup{colorlinks, citecolor=oucrimsonred, linkcolor=black, urlcolor=oucrimsonred}
\definecolor{cornellred}{rgb}{0.7, 0.11, 0.11}
\definecolor{navyblue}{rgb}{0.0, 0.0, 0.5}
\definecolor{amethyst}{rgb}{0.6, 0.4, 0.8}
\definecolor{yellow}{rgb}{1.0, 1.0, 0.0}
\definecolor{firebrick}{rgb}{0.7, 0.13, 0.13}
\definecolor{tangerineyellow}{rgb}{1.0, 0.8, 0.0}
\definecolor{deepfuchsia}{rgb}{0.76, 0.33, 0.76}
\definecolor{amber}{rgb}{1.0, 0.75, 0.0}
\definecolor{VioletRed4}{rgb}{0.55, 0.13, .32}
\definecolor{indiagreen}{rgb}{0.07, 0.53, 0.03}
\definecolor{VioletRed4}{rgb}{0.55, 0.13, .32}

\definecolor{oucrimsonred}{rgb}{0.6, 0.0, 0.0}
\newcommand\vertarrowbox[3][6ex]{%
  \begin{array}[t]{@{}c@{}} #2 \\
  \left\uparrow\vcenter{\hrule height #1}\right.\kern-\nulldelimiterspace\\
  \makebox[0pt]{\scriptsize#3}
  \end{array}%
}
\hypersetup{
     colorlinks   = true,
     citecolor    = violet,
     urlcolor     = violet,
     linkcolor    = violet}

\definecolor{verdechiaro}{rgb}{0.6,1,0.6}
\definecolor{giallochiaro}{rgb}{1,1,0.6}
\definecolor{bluscuro}{rgb}{0.15, 0.2, 0.9}
\definecolor{verdes}{rgb}{0.1, 0.5, 0.1}%
\definecolor{tangerineyellow}{rgb}{1.0, 0.8, 0.0}

\definecolor{americanrose}{rgb}{1.0, 0.01, 0.24}
\definecolor{cobalt}{rgb}{0.0, 0.28, 0.67}
\definecolor{brandeisblue}{rgb}{0.0, 0.44, 1.0}
\definecolor{mycolor}{rgb}{0.0, 0.0, 0.5}
\definecolor{oxfordblue}{rgb}{0.0, 0.13, 0.28}
\definecolor{azure}{rgb}{0.0, 0.5, 1.0}
\definecolor{turquoiseblue}{rgb}{0.0, 1.0, 0.94}
\newtcolorbox{mynewbox}[1]{colback=white!5!white,colframe=azure!75!black,fonttitle=\bfseries,title=#1}
\newtcolorbox{mybox}{colback=mycolor!5!white,colframe=azure!75!black}
\newtcolorbox{mynamedbox}[1]{colback=mycolor!5!white,colframe=azure!75!black,title=#1}
\definecolor{venetianred}{rgb}{0.78, 0.03, 0.08}
\newtcolorbox{mynamedbox1}[1]{colback=venetianred!5!white,colframe=venetianred!80!black,title=#1}
\newtcolorbox{mynamedbox2}[1]{colback=azure!5!white,colframe=azure!80!black,title=#1}

\definecolor{verdes}{rgb}{0.1, 0.5, 0.1}%
\definecolor{cornellred}{rgb}{0.7, 0.11, 0.11}

\definecolor{VioletRed4}{rgb}{0.55, 0.13, .32}

\definecolor{rossocorsa}{rgb}{0.83, 0.0, 0.0}

\title{Black Hole Photon Rings Saturate \\ the Quantum Chaos Bound}

\author[a,b]{D.~Giataganas\orcidlink{0000-0003-2003-3902},}
\author[c,d]{G.F.~Giudice\orcidlink{0000-0002-0247-4096},} 
\author[e]{A.~Kehagias\orcidlink{0000-0001-6080-6215},}
\author[c,f]{F.~Quevedo\orcidlink{0000-0002-7810-3662},}
\author[g,h]{A.~Riotto\orcidlink{0000-0001-6948-0856}}

\affiliation[a]{Department of Physics, National Sun Yat-Sen University, Kaohsiung 80424, Taiwan
}
\affiliation[b]{Physics Division, National Center for Theoretical Sciences, Taipei 10617, Taiwan}

\affiliation[c]{New York University Abu Dhabi, Saadiyat Island, Abu Dhabi, UAE}
\affiliation[d]{CERN, Theoretical Physics Department, 1211 Geneva, Switzerland}
\affiliation[e]{Physics Division, National Technical University of Athens, Athens 15780, Greece}
\affiliation[f]{DAMTP, University of Cambridge, Wilberforce Road, Cambridge CB2 0WA, UK}
\affiliation[g]{Department of Theoretical Physics and Gravitational Wave Science Center,  \\
24 quai E. Ansermet, CH-1211 Geneva 4, Switzerland}
\affiliation[h]{INFN, Sezione di Roma, Piazzale Aldo Moro 2, 00185 Rome, Italy}

\rightline{CERN-TH-2026-100}

\abstract{We study the quantum chaos bound in the photon ring region surrounding black holes. By evaluating the Lyapunov exponent associated with unstable null geodesics in a broad class of generalized Kerr geometries, as well as the temperature induced by a string probe, we show that the quantum chaos bound is exactly saturated on equatorial circular orbits of the photon ring. We confirm our result by deriving the same exponent from out-of-time-order correlators in the near ring region. As a byproduct, we show that the photon ring  saturation of the quantum chaos bound   implies the saturation of the Bekenstein bound on the rate of information emission from the ringdown phase through the quasi-normal modes in the eikonal limit. Our results extend the known correspondence between black hole thermodynamics and chaotic dynamics, highlighting the role of the photon ring as a probe of the fundamental limits on thermalization and information scrambling in black holes. }

\emailAdd{dimitrios.giataganas@gmail.com}
\emailAdd{gg3215@nyu.edu}
\emailAdd{kehagias@central.ntua.gr}
\emailAdd{fq2054@nyu.edu}
\emailAdd{Antonio.Riotto@unige.ch}
\begin{document}
\maketitle
\flushbottom

\section{Introduction}
It is often argued that the approach to thermalization is governed by chaotic
behavior. In classical systems, thermalization is commonly associated with
ergodicity and mixing. For generic initial conditions, time-averaged observables become representative of the accessible region of phase space, correlation functions decay, and the system effectively loses memory of its initial microscopic state. In many examples, this loss of memory is driven by chaos: nearby configurations in phase space separate exponentially fast. Specifically, if $\delta_0$ is the initial phase-space separation of two configurations, the separation at time $t$ is $\delta (t)=\delta_0 \, e^{\lambda t}$, where chaos is measured by the positive Lyapunov exponent $\lambda$.

In quantum mechanics, this classical picture must be reformulated. The overlap
between two quantum states remains constant under unitary time evolution, since
$\langle \psi_1 (t) | \psi_2 (t)\rangle =\langle \psi_1 (0) | \psi_2 (0)\rangle $. It is therefore more appropriate to
diagnose chaos through commutators of operators at different times. For two
generic Hermitian operators $W(t)$ and $Z(0)$, the relevant object is the
Out-of-Time Ordered Correlation function (OTOC) ~\cite{LarkinOvchinnikov1969}
\begin{equation}
C(t) = -Z^{-1} \text{Tr}\left(  e^{-\beta H} [W(t),Z(0)]^2 \right),
\label{larkin}
\end{equation}
where the thermal average refers to the temperature $T=(k_B \beta)^{-1}$. Since
the commutator is anti-Hermitian, $C(t)$ is real and positive. At early times,
the correlator in \eq{larkin} exhibits exponential growth,
$C(t)\propto e^{2\lambda_L t}$, allowing one to define a quantum analog of the
classical Lyapunov exponent.

Maldacena, Shenker, and Stanford proved, under suitable assumptions, that the
Lyapunov exponent characterizing the exponential growth of OTOCs in any thermal
quantum system satisfies~\cite{Maldacena:2015waa}
\begin{equation}\label{chaos}
\lambda_L\leq\frac{2\pi k_B T}{\hbar}.
\end{equation}
The bound (\ref{chaos}) establishes a fundamental speed limit on how fast a
thermal quantum system can scramble information \cite{Dvali:2013vxa}. Its quantum nature is manifest in the explicit factor of $\hbar$.

Black holes provide the canonical gravitational realization of maximal chaos.
When a black hole is perturbed by highly boosted quanta, the near-horizon
dynamics is captured by shock-wave scattering, and the corresponding eikonal
phase controls the growth of OTOCs in the dual thermal system
\cite{Shenker:2013pqa,Polchinski:2015cea}. In Einstein gravity the Lyapunov
exponent is fixed by the black hole surface gravity $\kappa_{\rm H}$,
\begin{equation}
\lambda_L =  \frac{\kappa_{\rm H}}{c} = \frac{2\pi k_B T_{\rm H}}{\hbar},
\label{boundsatur}
\end{equation}
where $T_{\rm H}$ is the Hawking temperature. Thus black holes saturate the
quantum chaos bound, in agreement with the expectation that they are the
fastest scramblers in nature~\cite{Hayden:2007cs,Sekino:2008he}.

A different but closely related structure in black hole physics is the photon
ring. It is formed by unstable null geodesics and controls both the optical
appearance of the black hole shadow and the eikonal limit of ringdown
perturbations. As a critical observational structure, the photon ring encodes
important information about black hole geometry and dynamics
\cite{Hadar:2022xag,Kapec:2022dvc,Giataganas:2024hil,Giataganas:2026ayt}.
Interest in photon rings has intensified following the image of the
supermassive black hole shadow in the galaxy $M87^*$ by the Event Horizon
Telescope collaboration~\cite{EventHorizonTelescope:2019dse}. Although current
observations do not yet resolve the photon ring unambiguously, future
space-based interferometry is expected to probe its structure with much higher
precision~\cite{Chael:2021rjo,Johnson:2019ljv}.

The aim of this work is to connect these two aspects of black hole physics from a new viewpoint. We probe the near-ring region with a finite tension string rather than a pointlike particle. A particle, or equivalently the eikonal limit of a field, samples the geometry only along a single worldline, and its instability is encoded in the Lyapunov exponent of the corresponding null geodesic.  In contrast, a string is an extended object with intrinsic worldsheet dynamics whose embedding coordinates $X^\mu(\tau,\sigma)$  satisfy the Nambu-Goto equations. Consequently, the induced worldsheet geometry can reveal structures that are invisible to a pointlike probe.

We focus on the equatorial near-ring region of Kerr. The appropriate local description is obtained by taking the Penrose limit around the equatorial circular null geodesic. In Brinkmann coordinates this limit gives a pp-wave geometry whose profile is determined by the tidal data of the Kerr photon orbit~\cite{blau,Fransen:2023eqj,Giataganas:2024hil}. We show that a string probe with a simple spacetime embedding in static gauge solves the Nambu-Goto equations in this background, yielding an induced metric on the string worldsheet that corresponds to a genuine two-dimensional Rindler geometry. The induced metric being Rindler means that observers using the worldsheet time see an Unruh temperature $T_{\rm ind}$ associated with the worldsheet Rindler horizon. Hence, the worldsheet theory is thermal with respect to Rindler time, provided one assumes the regular vacuum state across the worldsheet horizon.  The temperature relevant for the near-ring chaos bound is not the Hawking temperature $T_{\rm H}$ of the Kerr black hole, nor the temperature assigned to
a pointlike observer. It is the Unruh temperature associated with the Rindler horizon that appears on the worldsheet of the probe string. 

This observation provides the interpretation of the quantum chaos bound on the photon ring. It relates the four-dimensional Lyapunov exponent $\lambda$, which characterizes the instability of the photon ring, to a two-dimensional effective temperature $T_{\rm ind}$ on the string worldsheet.  We therefore obtain a near-ring version of the chaos bound, in which the role played by the black hole horizon instability in the standard near-horizon case is replaced by the photon ring instability characterized by $\lambda$ and the Hawking temperature by the effective Unruh temperature.

This is the main result of the present work. The equatorial photon ring of Kerr, described locally by its Penrose limit, induces a Rindler geometry on the worldsheet of an appropriate probe string. The corresponding Unruh temperature is determined by the photon ring Lyapunov exponent. In this sense, the photon ring is not only the locus of classical null geodesic instability. Instead, when probed by an extended string, it also gives rise to a thermal two-dimensional worldsheet theory that saturates the quantum chaos bound.

We will also comment on the relation between this worldsheet interpretation and the physics of the ringdown phase in black hole mergers. In the eikonal limit, where ringdown modes are controlled by photon ring geodesics, the same geometric data that determine the instability of null rays also determine the effective thermal scale on the probe string worldsheet. This provides a natural link between photon ring chaos, worldsheet thermality, and information emission during ringdown.

The paper is organized as follows. Section 2 contains a summary of the calculation of the Lyapunov exponents for generalized Kerr geometries.  In section 3 we introduce the probe string and derive the relevant Nambu--Goto equations in the near-ring pp-wave background. In section 4 we discuss the saturation of the quantum chaos bound. Section 5 explains the quantum nature of the chaos bound on photon rings in the shock-wave approximation, and section 6 illustrates its relation to the Bekenstein information  bound. Our conclusions are given in section 7. In Appendix A, we show that, when the instability is measured with respect to asymptotic coordinate time, the orbit-averaged Lyapunov spectrum in the Kerr photon shell is bounded below by the prograde equatorial orbit and bounded above by the retrograde equatorial orbit. Non-equatorial spherical photon orbits interpolate between these two limits. From now on, we will use natural units ($G_N=\hbar =c=k_B=1$) unless otherwise stated.

\section{The black hole Lyapunov exponents}
The first step to establish our result is the computation of the Lyapunov exponents associated with the instability of the photon ring. We consider a general stationary axisymmetric spacetime that admits a Killing--Yano tensor and possesses an analytic Penrose plane-wave limit \cite{Giataganas:2024hil}. This class includes, among others, Kerr, Kerr--Newman, and the static spherically symmetric case of the Schwarzschild black hole.
In Boyer--Lindquist coordinates  $(t,r,\theta,\phi)$, the metric takes the form 
\be
\dd s^2 = g_{tt}\,\dd t^2 +2g_{t\phi}\,\dd t\,\dd\phi +
g_{\phi\phi}\,\dd\phi^2+ g_{\theta\theta}\,\dd\theta^2  +g_{rr}\,\dd r^2
\label{metricapp}
\ee
\be
g_{tt} = \ff{a^2 s_\th^2-\D}{\Sigma},~~~
g_{t\phi}=\ff{a\prt{\D-K}s_\th^2}{\Sigma},~~~
g_{\phi\phi} =\ff{\left(K^2-a^2\D s_\th^2\right)s_\th^2}{\Sigma},~~~
g_{\theta\theta}= \Sigma ,~~g_{rr}= \ff{\Sigma}{\D},
\nonumber
\ee
\be
\Sigma  = r^2+a^2c_\th^2,\qquad
K = r^2+a^2, \qquad
s_\th=\sin\th,\qquad c_\th=\cos\th ,
\ee
which satisfies the identity
\be\label{eq:identityaa}
g_{t\phi}^2-g_{tt}g_{\phi\phi}=\Delta s_\th^2.
\ee
Here $a$ is the angular momentum parameter, $\D(r)$ is an arbitrary blackening factor that parametrizes the class of metrics under consideration. This spacetime admits two commuting Killing fields
\begin{equation}
\xi=\partial_t,\qquad \psi=\partial_\phi.
\end{equation}
For a circular orbit with constant angular velocity $\omega=\dd\phi/\dd t$, we define the associated helical Killing field
\begin{equation}
\chi := \xi+\omega\,\psi=\partial_t+\omega\,\partial_\phi.
\label{eq:helical-killing}
\end{equation}

The rotating black holes in this class are characterized by a photon shell, namely the spacetime region containing unstable bound null geodesics that neither escape to infinity nor cross the event horizon over extended times.  The photon shell is confined to the radial range $r_- \leq r \leq r_+$. Along the equatorial band $r_- \le r \le r_+$ at $\theta = \pi/2$, each point corresponds to a unique bound null geodesic. Away from the equator, for radii in the same interval, the trajectories oscillate in the polar direction between two turning angles $\theta_\pm$. Thus, the photon shell fills the spacetime region $r_- \le r \le r_+$, $\theta_- \le \theta \le \theta_+$, $0 \le \phi < 2\pi$. These bound null geodesics are inherently unstable: infinitesimal perturbations cause the photon either to fall into the black hole or to escape to infinity. The observable photon ring is generated by light rays that closely follow such nearly bound paths before eventually escaping.

To quantify this instability, consider two nearby geodesics, for instance one exactly bound at $r_\pm$, and another initially displaced by a small radial offset $\delta r_0$. According to the geodesic deviation equation, after a time $t$ their separation grows as
\begin{align}
\label{eq::Geodesic_Deviation}
\delta r(t) = e^{\lambda t}\delta r_0,
\end{align}
where a positive Lyapunov exponent $\lambda$ signals instability.

Restricting the geometry to the equatorial plane $\theta=\pi/2$, the non-vanishing components of the metric \eqref{metricapp} reduce to
\be
g_{tt} = \ff{a^2-\D}{r^2},~~~ g_{t\phi}=-a(g_{tt}+1), ~~~
g_{\phi\phi} = a^2(g_{tt}+2)+r^2,~~~ g_{rr}=\ff{r^2}{\D}.
\label{eqmeteq}
\ee
All metric components $g_{ij}$ are functions only of $r$ and satisfy the identity \eq{eq:identityaa}
\be
g_{t\phi}^2-g_{tt}g_{\phi\phi}=\Delta.
\label{eqidentity}
\ee
Using \eq{eqidentity}, we find that the conserved energy $E=-p_t$ and angular momentum $L=p_\phi$ satisfy
\begin{equation}\label{dotdot1}
\dot t=\frac{Eg_{\phi\phi}+Lg_{t\phi}}{\Delta},
\qquad
\dot\phi=-\frac{Eg_{t\phi}+Lg_{tt}}{\Delta},
\end{equation}
where dots denote derivatives with respect to the geodesic affine parameter $\sigma$. The null geodesic condition then yields the radial equation of motion in the compact form
\begin{equation} \label{calR1}
r^2\dot r^2=\cR(r),\qquad \cR(r):= E^2 g_{\phi\phi}+2EL\,g_{t\phi}+L^2 g_{tt}.
\end{equation}
Hence, $\cR(r)/r^2$ plays the role of an effective potential, while $\cR(r)$ fully encodes the radial dynamics.

For a circular null orbit at radius $r_0$ with angular velocity $\omega_0$, we have $\dot{\phi}=\omega_0 \dot{t}$ and the orbit is determined by
\begin{equation} \label{eomsR}
\mathcal R(r_0)=0,\qquad  \mathcal R'(r_0)=0.
\end{equation}
The orbital angular velocity follows from the first condition in \eq{eomsR}, and is given by the compact expression
\begin{eqnarray}\label{omega2}
\omega_{0}
&=&\frac{-g_{t\phi}(r_0)\pm \sqrt{\Delta(r_0)}}{g_{\phi\phi}(r_0)}.
\end{eqnarray}
The radial Lyapunov instability exponent with respect to coordinate time $t$ is then
\be \label{Lyap_definition}
\lambda^2=\frac{\cR''(r_0)}{2r_0^2\,\dot t_0^{\,2}},
\ee
where ${\dot t}_0 = \left.\dd t/\dd\sigma \right|_{r=r_0}$. This can be rewritten as
\begin{eqnarray} \label{lambdat}
\lambda^2=\frac{g_{tt}''(r_0)+2\omega_0 g_{t\phi}''(r_0)+\omega_0^2 g_{\phi\phi}''(r_0)}{2g_{rr}(r_0)}.
\end{eqnarray}
Here we have used the fact that \eq{eomsR} implies $E=\omega_0 L$ and it is easy to check that the right-hand side of \eq{lambdat} is always positive. This is the Lyapunov exponent governing the instability of equatorial null geodesics in the geometry \eq{metricapp}.

It is instructive to see the form that our expressions take in the two most representative cases within the class of metrics under consideration.

For the Schwarzschild black hole, the angular momentum parameter vanishes and the blackening factor is $\Delta(r)= r^2-2Mr$. The circular orbit conditions in Eqs.~\eqref{eomsR} reduce to
\be
\omega_0^2=\frac{r_0-2M}{r_0^3},\qquad \omega_0^2=\frac{M}{r_0^3},
\ee
which are solved by $r_0=3M$ and $\omega_0=\pm 1/(3\sqrt{3}\,M)$. In this case, the sign of the angular velocity is physically irrelevant, since the two directions are equivalent. Substituting the radius and angular frequency into \eqref{lambdat}, the Lyapunov exponent reduces to $\lambda=1/(3\sqrt{3}M)$.

For the Kerr black hole, the blackening factor is $\Delta(r)= r^2-2Mr+a^2$. From Eqs.~\eqref{eomsR}, restricted to the equatorial plane, one obtains
\begin{equation}
r_{0}^2-3Mr_{0}\pm 2a\sqrt{Mr_{0}}=0,
\label{eq:rph-equation}
\end{equation}
which yields two photon ring radii, in contrast to the static case. In closed form, the solutions are
\begin{equation}\label{radiiph}
r_\pm = 2M\left[1+\cos\left(\frac{2}{3}\arccos\left(\pm\frac{a}{M}\right)\right)\right].
\end{equation}
The corresponding angular frequencies are
\be
\omega_{\pm}=\frac{\sqrt{M}}{a\sqrt{M}\mp (r_{\pm})^{3/2}},
\label{eq:Omega}
\ee
where the sign distinguishes the two directions of rotation. Note that for $a=0$ one finds $r_\pm=3M$, as expected, while at extremality $a=M$ one finds $r_{-}=M$ for the prograde orbit and $r_{+}=4M$ for the retrograde one. The Kerr Lyapunov exponents are obtained from \eq{lambdat} and take the  form~\cite{Cardoso:2008bp,Deich:2023oox}

\begin{equation}
\label{lambdakerr}
\lambda_{\pm}=
\frac{\sqrt{3\,\Delta_{\pm}}}{r_{\pm}\,\sqrt{a^2+3r_{\pm}^2}},
\qquad
\Delta_\pm=\Delta(r_\pm)~.
\end{equation}

\section{The induced temperature}
After having calculated the Lyapunov exponents of the photon rings, the next  crucial step toward establishing the saturation of the quantum chaos bound  is to identify the correct thermodynamical counterpart and, specifically, the  temperature associated with the photon ring. To this aim, we need to take an intermediate step, zoom in on the photon ring, and use a probe. For this purpose, a string   provides a more refined diagnostic of the Kerr geometry than a pointlike particle. The latter, or equivalently the eikonal limit of a field, samples the geometry only through a single worldline. In this limit the dynamics is governed by the null geodesic equations, and near the photon ring the instability is encoded in the geodesic Lyapunov exponent. By contrast, a finite-tension string is an extended object with intrinsic worldsheet dynamics.

The use of an extended probe is not ad hoc.  Over the years, strings in strong gravitational fields have been studied (see for instance, \cite{Horowitz:1990sr, Dodelson:2015toa}). In holographic non-equilibrium systems, probe branes and strings routinely generate effective horizons on their induced worldvolume geometries, and the corresponding  Hawking temperatures of such  horizons govern the fluctuations of the probe degrees of freedom.  The logic is simple  but powerful.  The external driver injects energy into a localized probe sector, while the ambient geometry acts as a heat bath and energy sink.  In the probe limit, the backreaction on the bulk geometry is negligible, but the induced metric on the probe is modified non-trivially.  The probe degrees of freedom then propagate not directly in the original bulk metric, but in an effective worldvolume geometry.  When this geometry develops a horizon, its surface gravity defines an effective temperature, generally different from the temperature of the ambient black-hole background 
\cite{Nakamura:2013yqa,Sonner:2012if,Giataganas:2013hwa}\footnote{We thank J. Sonner for discussions about this point.}.

This mechanism has already been used in the literature and motivates our construction.  In driven probe-brane systems, the induced worldvolume horizon carries an induced temperature $T_{\rm ind}$, and this temperature controls fluctuation-dissipation type relations even far from equilibrium \cite{Nakamura:2013yqa}.  Similarly, when an electric field is applied to a probe brane, the induced metric is modified so that an emergent horizon appears. Hawking radiation from this horizon propagates to the boundary and is interpreted as thermal current noise in the dual quantum system \cite{Sonner:2012if}.  Thus,  the appearance of a temperature on an extended probe is not a phenomenological assumption, but a geometric consequence of the induced worldvolume horizon. The lesson from these examples is that extended probes can couple to structures that are absent in a purely worldline description.  Their induced worldvolume geometry can develop horizons whose surface gravity defines an induced temperature, and their worldsheet couplings can be sensitive to global or topological data of the background.

Our probe string plays precisely this role for the photon ring.  A
pointlike particle samples only a single null trajectory and therefore
captures only the geodesic Lyapunov exponent. By contrast, a finite-tension string carries an intrinsic two-dimensional worldsheet theory governed by the induced metric.  In the near-ring Penrose limit, this induced metric becomes Rindler, with a genuine worldsheet horizon.  The temperature entering the chaos bound is then not assigned by hand to a photon orbit, nor identified with the Hawking temperature of the four-dimensional black hole.  It is the Unruh temperature of the induced Rindler horizon on the probe string worldsheet.  In this sense, the probe string converts the classical four-dimensional instability of the photon ring into a two-dimensional thermal problem.

To make this statement precise, we describe the string by its spacetime position $X^\mu(\tau,\sigma)$, viewed as a set of dynamical fields living on the two-dimensional worldsheet. These fields satisfy the Nambu--Goto equations of motion. We will apply this description to a probe string localized in the near-ring region of the Kerr geometry. The construction  extends straightforwardly to the generalized class of black hole geometries considered throughout this work. The photon ring is controlled by unstable null geodesics, and the Penrose limit around such a geodesic isolates the local geometry seen by an observer following that orbit. In Brinkmann coordinates, this local geometry takes the pp--wave form \cite{blau,Fransen:2023eqj,Giataganas:2024hil}
\begin{equation}
\dd s^2=2\,\dd u\, \dd v+A_{ij}(u)x^i x^j\dd u^2+\dd x_1^2+\dd x_2^2 ,~~~~i,j=1,2
\label{pp-Aij}
\end{equation}
where 
\begin{equation}
A_{11}=-A_{22}=\frac{12M r_0^3\Delta}{(r_0-M)^2\Sigma^5}
\left[5\left(r_0^2-a^2\cos^2\theta\right)^2-4r_0^4
\right],
\label{A11-A22-general}
\end{equation}
\begin{equation}
A_{12}=A_{21}=a\cos\theta\,
\frac{12M r_0^2\Delta}{(r_0-M)^2\Sigma^5}\left[5\left(r_0^2-a^2\cos^2\theta\right)^2-4a^4\cos^4\theta\right].
\label{A12-general}
\end{equation}
For the equatorial orbit, $\theta=\pi/2$, the pp-wave metric (\ref{pp-Aij}) that solves the vacuum Einstein equations turns out to be
\begin{equation}
\dd s^2=2\,\dd u\,\dd v+A_{11}\Big( x_1^2-x_2^2\Big)\dd u^2+\dd x_1^2+\dd x_2^2 ,
\label{a11}
\end{equation}
where, for example, using  the case of Kerr functions $\Sigma_0\equiv \Sigma (r_0)=r_0^2$ and $\Delta_0\equiv \Delta(r_0)=r_0^2-2Mr_0+a^2$, we find
\begin{equation}
A_{11}=\frac{12M\Delta_0}{(r_0-M)^2r_0^3}.
    \label{a11=a22}
\end{equation}
Let us recall that  the equatorial Kerr
null radial potential is given by \eq{calR1}
\begin{equation}
\mathcal{R}(r)=\left[E(r^2+a^2)-aL\right]^2-\Delta(L-aE)^2.
\label{kerr-radial-pot}
\end{equation}
Then it is straightforward to verify that 
\begin{equation}
A_{11}=\frac{{\cal R}''(r_0)}{2r_0^4},
\label{A11-lambda-u}
\end{equation}
so that $A_{11}$ is written in terms of the Lyapunov exponent $\lambda$ as 
\begin{eqnarray}
A_{11}=\dot{t}_0^2 \, \lambda^2,
\label{A11-lambda}
\end{eqnarray}
where $\dot{t}_0$ denotes the derivative with respect to the geodesic affine parameter, evaluated at the photon ring. We would now like to probe the near-ring  Kerr geometry with a probe string. The worldsheet coordinates are denoted by $\xi^a=(\tau,\sigma)$, with $a,b=0,1$. The embedding of the string worldsheet into spacetime is
\begin{equation}
X^\mu=X^\mu(\xi^a),\qquad \mu,\nu=0,1,2,3 .
\end{equation}
Then the dynamics of the probe string is described by the Nambu--Goto action 
\begin{equation}
S_{\rm NG} = -T_{\rm s} \int \dd^2\xi\,\sqrt{-\det G_{ab}},\label{NG}
\end{equation}
where $T_{\rm s}$ is the string tension and $G_{ab}$ is the induced metric. Using reparametrization invariance, we can choose for the probe string in the near-ring Kerr geometry given by the pp-wave metric (\ref{a11}), the static gauge  $t=\tau,$ $x_2=\sigma,$ 
which is equivalently written as 
\begin{eqnarray}
u=\frac{\tau}{\dot t_0}, \qquad x_2=\sigma.
\end{eqnarray}
Then, the gauge-fixed Nambu--Goto Lagrangian density for the rest of the fields $v(\tau,\sigma)$ and $x_1(\tau,\sigma)$ turns out to be
\bea
\mathcal L_{\rm NG}&=&-T_{\rm s}\left\{\left(\frac{\partial_\sigma v}{\dot t_0}+\partial_\tau x_1\,  \partial_\sigma x_1
\right)^2\right.\nonumber \\ &-&\left. \left[\frac{A_{11}(x_1^2-\sigma^2)}{\dot t_0^2} +\frac{2\partial_\tau v}{\dot t_0}+(\partial_\tau  x_1)^2\right]
\Big[1+(\partial_\sigma x_1)^2\Big]\right\}^{1/2}.
\eea
A solution for $v(\tau,\sigma)$ and $x_1(\tau,\sigma)$ is 
\begin{eqnarray}
v(\tau,\sigma)=0, \qquad x_1(\tau,\sigma)=0,
\end{eqnarray}
so that the induced worldsheet metric on the probe string becomes
\be
\dd s_{\mathrm{ws}}^2=-\frac{A_{11}}{\dot t_0^2}\,\sigma^2\,\dd \tau^2 + \dd \sigma^2 .    \label{ds1}
\end{equation}
It is clear now from \eq{ds1} that  a genuine  Rindler geometry on the worldsheet is revealed. 
In particular, by writing the metric (\ref{ds1}) as 
\begin{equation}
\dd s_{\mathrm{ws}}^2= -\kappa_{\rm ind}^2\,  \sigma^2\,\dd \tau^2 + \dd \sigma^2 ,
\label{ds2}
\end{equation}
we see that 
 there is a horizon at $\sigma =0$, which is the position of the photon ring,  with corresponding surface gravity
\begin{equation}
\kappa_{\rm ind} = \frac{\sqrt{A_{11}}}{\dot t_0}. \label{k1}
\end{equation}
Moving to Euclidean worldsheet time $(\tau=-i \tau_{\rm E})$,  the metric (\ref{ds2}) becomes
\begin{equation}
 \dd s_{\mathrm{Ews}}^2= \kappa_{\rm ind}^2\,  \sigma^2\,\dd \tau_{\rm E}^2 + \dd \sigma^2 . \label{dd}
\end{equation}
This is the  metric of a two-dimensional plane in polar coordinates, with no conical singularities provided that the angular variable $\kappa_{\rm ind} \tau_{\rm E}$ has period $2\pi$. Therefore, regularity at $\sigma=0$ requires the Euclidean time to be periodic under
\begin{equation}
\tau_{\rm E} \to \tau_{\rm E} + \frac{2\pi}{\kappa_{\rm ind}} .
\end{equation}
Comparing this with the standard Euclidean-time periodicity of a thermal system, $\tau_{\rm E} \to \tau_{\rm E} + \beta$, we can define an induced temperature $T_{\rm ind} =1/\beta$ given by  
\begin{equation}
T_{\rm ind} = \frac{\hbar\, \kappa_{\rm ind}}{2\pi  k_B}.
\label{Teff}
\end{equation}
Note that the worldsheet temperature $T_{\rm ind}$  should not be confused with the Hawking temperature $T_{\rm H}$ of the Kerr black hole. Instead, $T_{\rm ind}$ is the Unruh   temperature associated with thermal effects on the worldsheet of the probe string. 

We note that a similar Rindler structure, characterized by the same  temperature, arises in the effective metric describing the near–photon-ring dynamics \cite{Giataganas:2024hil,Raffaelli:2021gzh}. However, as we have emphasized before, it is the string nature of the probe that allows us to make a proper identification of the chaos bound.

\section{The saturation of the quantum chaos bound}
A direct comparison of Eqs.~(\ref{A11-lambda}), (\ref{k1}) and (\ref{Teff}) gives
\begin{equation}
\lambda=\kappa_{\rm ind}=\frac{2\pi k_B T_{\rm ind}}{\hbar}. \label{lambda-T}
\end{equation}
Thus, the Lyapunov exponent governing the near-ring instability is equal to the surface gravity of the induced Rindler worldsheet metric. Consequently, the thermal worldsheet dynamics of the probe string saturates the quantum chaos bound in \eq{chaos}.

This shows that, in the near-ring Kerr geometry, the effective worldsheet temperature is directly tied to the classical instability of the photon ring dynamics,  elucidating how the connection between chaos and thermodynamics is encoded in the nature of the induced Rindler metric. 

To expand on the physical meaning of our result, let us first recall the interpretation of the quantum chaos bound for a black hole characterized by a Hawking temperature $T_H$. By dropping a probe mass into it, the black hole is driven out of thermal equilibrium. The system relaxes back to equilibrium through a chaotic process with a rate characterized by the Lyapunov exponent $\lambda = 2 \pi T_H$. This establishes the saturation of the quantum chaos bound for black holes.

Consider now the photon ring. Its null orbits are unstable and, with a pure gravity calculation that does not use any quantum or thermodynamical properties, we have computed the corresponding Lyapunov exponents, showing that they are equal to the surface gravity. Next, we considered the effect of dropping a string probe into the photon ring. This induces a local excitation on the corresponding worldsheet, which is interpreted as a quantum temperature $T_{\rm ind}$. The relaxation of these excitations back to thermal equilibrium is governed by the previously-computed Lyapunov exponents, which turn out to saturate the quantum chaos bound.

It is useful to apply our result to the two most representative cases of black holes. For Schwarzschild, $\keff^2$ reduces to  
\be
\keff^2=\frac{1}{27M^2},
\ee
which reproduces the known result~\cite{Raffaelli:2021gzh} and shows that the quantum chaos bound is saturated at $\Teff =(4/3\sqrt{3})T_H$.

\begin{figure*}[t!]
\centering
  \includegraphics[width=0.8\textwidth]{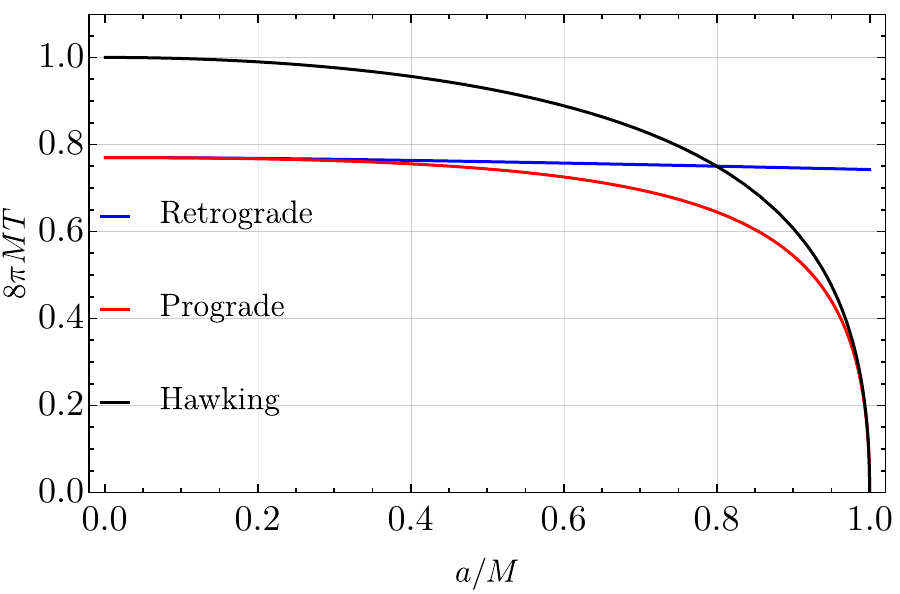}
  \caption{The induced temperatures on the prograde and retrograde equatorial circular orbits, as well as the Hawking temperature, for a Kerr black hole. All temperatures are expressed in units of $1/(8\pi M)$ and plotted as functions of the angular momentum parameter $a/M$.}
  \label{fig:1}
\end{figure*}
For a Kerr black hole, the corresponding surface gravity is 
\be
\keff^2 =
\frac{3\Delta(r_0)}{r_0^2(a^2+3r_0^2)},
\label{c}
\ee
which agrees with \eq{lambdakerr}. In \fig{fig:1} we plot the induced temperatures on the prograde and retrograde equatorial circular orbits, as well as the Hawking temperature, normalized in units of $1/(8\pi M)$. In the limit $a\to 0$, we recover the Schwarzschild result. In the near-extremal limit $a \to M$, the induced temperature $T_{\rm ind}$ for the prograde orbit tends towards the Hawking temperature of the Kerr black hole. This is not surprising, since in the extremal case the prograde orbit lies at the degenerate horizon $r_-=M$, and the angular frequency coincides with the orbital frequency of the black hole. At $a=M$, both $T_{\rm ind}$ on the prograde orbit and the Hawking temperature vanish. On the other hand, the induced temperature on the retrograde orbit is fairly insensitive to angular momentum, as shown in \fig{fig:1}. 
In Appendix A we show that the Lyapunov exponent across the Kerr photon shell is minimized by the equatorial prograde orbit, which therefore determines the relevant timescale for the chaotic process.

\section{On the quantum nature of the chaos bound}

We make here some important remarks about the quantum nature of the bound saturation, which might seem surprising since the Lyapunov exponent is purely classical. The point is that, although the Lyapunov exponent is classical, the induced temperature is not. We refer to the bound as quantum precisely because the induced temperature is ${\cal O}(\hbar)$. The chaotic behavior is experienced by states that are produced quantum mechanically and measured by the local Rindler observer as a thermal spectrum.  The situation is fully analogous to the quantum chaotic behavior observed near the horizon and experienced by the quantum Hawking radiation~\cite{Shenker:2013pqa,Polchinski:2015cea}. There, the Lyapunov exponent is purely classical and sets the timescale, as seen by an asymptotic observer, over which an outgoing Hawking particle escapes from the near-horizon region after the black hole is perturbed by an ultrarelativistic particle whose gravitational field is described by a shock wave. The quantum connection to the chaos bound is obtained by expressing the Lyapunov coefficient in terms of the  ${\cal O}(\hbar)$ Hawking temperature. Let us further elaborate on this analogy.

In the previous section, we have shown that the induced two-dimensional geometry governing near-critical photon ring geodesics takes the local Rindler form
\begin{equation}
\dd s_{(2)}^2= -\kappa_{\rm ind}^2 \rho^2 \dd t^2+\dd \rho^2 , \label{near-ring}
\end{equation}
where $t$ is the asymptotic time coordinate adapted to the circular orbit and $\rho$ is the proper radial distance from the null surface of the induced geometry. The effective surface gravity $\kappa_{\rm ind}$ is equal to the coordinate-time Lyapunov exponent of the unstable circular null orbit,   $\lambda=\kappa_{\rm ind}$. This equality is purely geometrical. We now explain how the same exponent arises from a local gravitational eikonal OTOC in the near-ring region.

Starting from Eq.~\eqref{near-ring}, let us introduce local Rindler null
coordinates
\begin{equation}
U=-\frac{1}{\sqrt{2}}\,\rho\, e^{-\kappa_{\rm ind}t},\qquad V=\frac{1}{\sqrt{2}}\,\rho \, e^{\kappa_{\rm ind}t}. \label{rindler-null-coor}
\end{equation}
Then, 
the two-dimensional metric becomes locally
\begin{equation}
 \dd s_{(2)}^2= -2\, \dd U\,\dd V .
\end{equation}
The essential property is that translations in Rindler time act as boosts:
\begin{equation}
t\rightarrow t+\Delta t  \quad\Longrightarrow\quad    U\rightarrow e^{-\kappa_{\rm ind}\Delta t}\, U,\qquad    V\rightarrow e^{\kappa_{\rm ind}\Delta t}\,V.
\label{rindler-boost}
\end{equation}
Therefore, an excitation inserted at an early Rindler time is exponentially boosted relative to a later excitation. This exponential boost is the kinematic origin of the Lyapunov growth.

Let $H_R$ denote the Hamiltonian generating translations in the near-ring
Rindler time $t$. $H_R$  can be  written as   $H_R=\kappa_{\rm ind}K,$ 
where $K$ is the boost charge, generating translations in the dimensionless
Rindler time $\eta=\kappa_{\rm ind}t$. We define the local thermal density matrix  \cite{Bisognano:1975ih,Bisognano:1976za}
\begin{equation}
\varrho_{\rm eff}= \frac{e^{-\beta_{\rm eff}H_R}}{\mathrm{Tr} e^{-\beta_{\rm eff}H_R}}, \qquad \beta_{\rm eff}=\frac{1}{k_BT_{\rm ind}}=\frac{2\pi}{\hbar \kappa_{\rm ind}}, \label{rho-eff}
\end{equation}
which is  naturally
associated with the Rindler patch of the induced near-ring geometry \cite{Crispino:2007eb}. 

It is known that the photon near-ring geometry in the eikonal approximation is described by the Penrose limit of background Kerr geometry \cite{blau,Fransen:2023eqj,Perrone:2025zhy,Giataganas:2024hil}. Perturbations in this  limit are generated by the Hertz potential $\Phi(x^\mu)$. We may choose now two near-ring operators $W$ and $Z$, which should be understood as gravitationally dressed operators that generate perturbations localized in the near-photon ring region.  Thus, one may write the following expressions 
\be
W = \int \dd U\,\dd^{2}x_\perp\, f_W(U,x_\perp)\, \Phi(U,x_\perp),~~~~ 
Z =\int \dd V\,\dd^{2}x_\perp\,  f_Z(V,x_\perp)\, \Phi(V,x_\perp),
\label{WZ-operator}
\ee
where $x_\perp$ denotes the coordinates perpendicular to the $U,V$ coordinates, and
$f_Z,f_W$ are wavepacket profiles.
The operator $W(t)$ is evolved with the local Rindler Hamiltonian,
\begin{equation}
W(t)= e^{iH_R t/\hbar} \, W(0) \, e^{-iH_R t/\hbar},
\end{equation}
and we define the OTOC by
\begin{equation}
F(t)=\mathrm{Tr}\!\left[\varrho_{\rm eff}W^\dagger(t) Z^\dagger(0) W(t)Z(0)\right].
\label{otoc-def}
\end{equation}
Equivalently, one may use the squared commutator
\begin{equation}
C(t)= -\mathrm{Tr}\!\left[\varrho_{\rm eff}  [W(t),Z(0)]^2    \right].
\label{sq-com}
\end{equation}
For chaotic systems, these quantities have an intermediate-time regime in which
\begin{equation}
F_{\rm d}-F(t)\sim \epsilon \,e^{\lambda_L t}, \qquad C(t)\sim \epsilon' \,e^{2\lambda_L t},
\label{Fd-Ft}
\end{equation}
where $F_{\rm d}$ is the disconnected or factorized value of the OTOC \cite{Shenker:2013pqa,Hashimoto:2017oit,Maldacena:2015waa}, $\lambda_L$ is the corresponding Lyapunov coefficient\footnote{Note that we distinguish between the geometric Lyapunov coefficient, computed before and denoted by $\lambda$, and the quantum Lyapunov coefficient defined here as $\lambda_L$. In the following we demonstrate that $\lambda=\lambda_L$.} and $\epsilon, \epsilon'$ are constants.

The OTOC~\eqref{otoc-def} may be approximated by a scattering amplitude in the near-ring Rindler region. This is the same physical mechanism that appears in the black hole shock-wave derivation of maximal chaos, where an early operator creates a highly boosted excitation, which gravitationally scatters a later excitation. The OTOC measures the sensitivity of the later excitation to the earlier perturbation \cite{Roberts:2014isa,Shenker:2013pqa,Maldacena:2015waa}.

Let us now consider one null wavepacket moving in one null direction with  momentum
\begin{equation}
p^{(1)}=p^{(1)}_a dx^a = p_U(t)\,\dd U ,\qquad p^{(1)}\cdot p^{(1)}=0,
\end{equation}
and let the other wavepacket move in the opposite null direction with covariant
momentum
\begin{equation}
p^{(2)}= p^{(2)}_a dx^a  = p_V\,\dd V , \qquad p^{(2)}\cdot p^{(2)}=0.
\end{equation}
We assume that the operator $W(t)$ creates the early excitation moving along one null branch, with momentum $p_U$, and  $Z(0)$ creates the later excitation moving along the opposite null branch, with momentum $p_V$. 
Then, the center-of-mass energy squared of the two near-ring wavepackets turns out to be
\begin{equation}
s(t)= -\left(p^{(1)}+p^{(2)}\right)^2= 2\, p_U(t)\, p_V .
\end{equation}
Rindler time translations act as boosts. Hence, an excitation inserted at an early Rindler time is boosted according to
\be
p_U(t)=p_U(0)\, e^{\kappa_{\rm ind}t},
\qquad
s(t)=2\, p_U (0) p_V\, e^{\kappa_{\rm ind}t}.
\ee
In the eikonal regime, the gravitational interaction is approximately elastic and diagonal in impact-parameter space. The only effect of the scattering is, therefore, to multiply the two-particle wavefunction by the phase
\begin{equation}
S_{\rm eik}(t,b) = e^{i\delta_{\rm eik}(t,b)}, \label{eik-S}
\end{equation}
where $b$ is the transverse impact parameter.  The OTOC may then be understood as an overlap between two states obtained by applying the same operators in different orders. Indeed,  one may purify the thermal density matrix $\varrho_{\rm eff}$  by introducing a thermofield-double state $|\beta_{\rm eff}\rangle$ in a doubled Hilbert space, satisfying \cite{Takahashi:1996zn,Israel:1976ur,Maldacena:2001kr}
\begin{equation}
\mathrm{Tr}\!\left[\varrho_{\rm eff}\mathcal O\right]=\langle\beta_{\rm eff}|\mathcal O|\beta_{\rm eff}\rangle .
\end{equation}
Then we may define the two states
\begin{equation}
|\psi_1(t)\rangle=W(t)Z(0)|\beta_{\rm eff}\rangle, \qquad |\psi_2(t)\rangle=Z(0)W(t)|\beta_{\rm eff}\rangle ,
\end{equation}
such that
\begin{equation}
F(t)=\langle \varrho_{\rm eff} W^\dagger(t)Z^\dagger(0)W(t)Z(0)\rangle=\langle\psi_2(t)|\psi_1(t)\rangle .
\end{equation}

In the absence of interactions, the two states coincide, up to trivial phases. In the gravitational eikonal regime, the two orderings differ because the earlier excitation creates a shock wave through which the later excitation propagates. Thus, one state is related to the other by the eikonal scattering operator,  $|\psi_1\rangle=S_{\rm eik}|\psi_2\rangle ,$  and therefore
\begin{eqnarray}
F(t)=\langle\psi_2|S_{\rm eik}|\psi_2\rangle .
\label{eq:psi2-psi1}
\end{eqnarray}

We now expand the two-particle state in a basis of null momenta and impact
parameter,
\begin{eqnarray}
|\psi_2\rangle=\int \dd p_U\,\dd p_V\,\dd b\,\Psi_W(p_U,b)\Psi_Z(p_V,b)|p_U,p_V,b\rangle .
\label{eq:psi-2}
\end{eqnarray}
In the eikonal approximation, the scattering is elastic and diagonal in this
basis,
\begin{eqnarray}
S_{\rm eik}|p_U,p_V,b\rangle=e^{i\delta_{\rm eik}(t,b)}|p_U,p_V,b\rangle ,
\end{eqnarray}
so that the OTOC reduces to an average of the eikonal phase over the wavepacket profiles.
Indeed, substituting then \eqref{eq:psi-2}  into \eqref{eq:psi2-psi1}, we find
\begin{eqnarray}
F(t)=\int \dd p_U\,\dd p_V\,\dd b\,|\Psi_W(p_U,b)|^2|\Psi_Z(p_V,b)|^2 e^{i\delta_{\rm eik}(t,b)} .
\label{otoc-eik-intl}
\end{eqnarray}
What remains is to calculate the phase shift $\delta_{\rm eik}$, which we will do in the shock-wave approximation \cite{Dray:1984ha,Camanho:2014apa}. 

The early excitation may be described as a gravitational shock wave. In a local near-ring patch, choose coordinates such that the unperturbed
metric is
\begin{equation}
\dd s^2 = -2\, \dd U\,\dd V+H(x_\perp)\,\dd U^2+ \dd x_\perp^2 .
\end{equation}
The function $H(x_\perp)$ satisfies 
\begin{eqnarray}
\nabla_\perp^2 H=0,
\end{eqnarray}
where $\nabla_\perp^2$ is the Laplacian in the transverse coordinate $x_\perp$. A null particle moving along $U=0$, with momentum $p_U$, produces a shock
geometry of the form \cite{Araneda:2022lgu,Kehagias:2024sgh}
\begin{equation}
\dd s^2 =-2\,\dd U\,\dd V+H(x_\perp)\,\dd U^2+ \dd x_\perp^2 +  \Phi(x_\perp)\,\delta(U)\,\dd U^2 . \label{eq:shock-metric}
\end{equation}
The only nonzero curvature component is proportional to the transverse Laplacian of the shock profile, 
\begin{equation}
R_{UU}=-\frac12 \nabla_\perp^2\Phi(x_\perp)\,\delta(U).
\end{equation}
The stress tensor of the null particle is
\begin{equation}
T_{UU}= p_U\,\delta(U)\,\delta^{(2)}(x_\perp),
\end{equation}
and Einstein's equation therefore reduces to 
\begin{equation}
\nabla_\perp^2\Phi(x_\perp)=-16\pi G_N\,p_U\, \delta^{(2)}(x_\perp).
\label{shock-phi}
\end{equation}

Let $G_\perp(x_\perp,x'_\perp)$ be the Green function satisfying
\begin{equation}
\nabla_\perp^2 G_\perp(x_\perp,x'_\perp) = \delta_\perp(x_\perp-x'_\perp),
\end{equation}
which for the flat 2D transverse space is given by
\begin{eqnarray}
G_\perp(x_\perp,x'_\perp)=\frac{1}{2\pi} \log|x_\perp-x'_\perp|. 
\end{eqnarray}
Then, from \eqref{shock-phi} we find that
\begin{equation}
\Phi(x_\perp) =-16\pi G_N\,p_U \, G_\perp(x_\perp,0).
\end{equation}
Taking the early excitation to be given by the boosted momentum $p_U=p_U(0)\, e^{\kappa_{\rm ind}t},$  we obtain
\begin{equation}
\Phi(t,x_\perp) = -16\pi G_N\,p_U(0) \, e^{\kappa_{\rm ind}t} \, G_\perp(x_\perp,0). \label{b-shock-prof}
\end{equation}
A probe crossing the shock at $U=0$ experiences a discontinuity in the null coordinate $V$, the Shapiro time delay. With the metric of \eq{eq:shock-metric}, the shift is
\begin{equation}
\Delta V(t,x_\perp) =\frac12 \Phi(t,x_\perp), \label{V-shift}
\end{equation}
and thus,
\begin{equation}
\Delta V(t,x_\perp) = -8\pi G_N\,p_U(0)\, e^{\kappa_{\rm ind}t}\, G_\perp(x_\perp,0).
\label{b-V-shift}
\end{equation}
A wavepacket with momentum $p_V$ crossing the shock acquires the phase $\exp\!\left[-ip_V\Delta V(t,x_\perp)\right]$
 and therefore the eikonal phase turns out to be
\begin{equation}
\delta_{\rm eik}(t,x_\perp)=-p_V\Delta V(t,x_\perp)=8\pi G_N\,p_U(0) \, p_V(0)\,e^{\kappa_{\rm ind}t}\,G_\perp(x_\perp,0).\label{eik-phase}
\end{equation}
Substituting Eq.~\eqref{eik-phase} into the eikonal expression for the OTOC, Eq.~\eqref{otoc-eik-intl}, gives
\begin{equation}
F(t)=\int \dd p_U\,\dd p_V\,\dd b\,|\Psi_W|^2|\Psi_Z|^2\exp\!\left[i\,8\pi G_N\,p_U(0) \, p_V(0)\, e^{\kappa_{\rm ind}t}\,  G_\perp(b)\right]. \label{ft-ft}
\end{equation}
At early times and in the eikonal regime, the phase is small,
\begin{equation}
G_N\,p_U(0)\, p_V(0) \,e^{\kappa_{\rm ind}t}\, \ll1,
\end{equation}
so that we have 
\begin{equation}
F(t)=F_{\rm d}+8\pi i\, G_N\,e^{\kappa_{\rm ind}t} \int \dd p_U\,\dd p_V\,\dd b\, |\Psi_W|^2|\Psi_Z|^2\, p_U(0)\, p_V(0)\, G_\perp(b)+\cdots ,
\end{equation}
where 
\begin{equation}
F_{\rm d}= \int \dd p_U\,\dd p_V\,\dd b\, |\Psi_W|^2|\Psi_Z|^2.\label{toc-phase-int-0}
\end{equation}
After the appropriate choice of wavepackets and operator ordering, the leading connected correction has the form
\begin{equation}
F_{\rm d}-F(t)= \epsilon\,e^{\kappa_{\rm ind}t}+  \cdots ;
\label{otoc-lyu}
\end{equation}
Eq.~\eqref{otoc-lyu} is the desired local OTOC growth law. Comparing
with the standard definition of Eq.~\eqref{Fd-Ft},
we identify
\begin{equation}
\lambda_L=\kappa_{\rm ind}=\lambda. \label{lam-ka}
\end{equation}
Thus, the local near-ring eikonal OTOC has the same exponent as the Rindler
surface gravity of the induced photon-ring geometry.

In the next section, we will provide  evidence for the relation between the saturation of the quantum chaos bound and the physics of the ringdown. Indeed, it is in this phase that the information about the physics of the photon ring is transmitted to an asymptotically far  observer through the emission of gravitational  waves. As we will show, the saturation of the quantum chaos bound is expressible in terms of the saturation of the Bekenstein bound on the information rate.

\section{The saturation of the Bekenstein information bound} 
The rate at which a physical system can convey information depends on its geometry, the nature of the information carrier, and the dynamics governing the emission. Nevertheless, independently of these details, there exists a universal upper bound on the information transfer rate. This result traces back to an argument by  Bekenstein regarding the maximal information content of a static system~\cite{Bekenstein:1981zz}. By combining the generalized second law of thermodynamics with causality, Bekenstein derived a fundamental bound on the  average rate of information emission achievable by any physical system
\be
\label{info}
\frac{\dd I}{\dd t}\leq\frac{\pi E}{\hbar \ln 2},
\ee
where $E$ is the signal energy in the receiver’s frame. Since entropy can be viewed
as a measure  of the lack of information about the internal state of the system, any decrease in information amounts to an increase in entropy. Therefore,
\eq{info} implies an  upper bound on the rate of change of entropy
\be
\frac{\Delta S}{\tau}\leq \frac{\pi k_B \Delta E}{\hbar}.
\ee
Here  $\tau$ is the relaxation time required for the perturbed system to return to an equilibrium state. Using the second law of thermodynamics $\Delta S \geq \Delta E/T$, one obtains the bound~\cite{Hod:2006jw} 
\be
\tau \geq \frac{\hbar}{\pi k_B T},
\label{bekhod}
\ee
where $T$ is the temperature of the system.

During the ringdown, the final stage of the merger between two black holes, Quasi-Normal Modes (QNMs) characterize the response of the remnant black hole to perturbations, with frequencies that depend solely on its mass, spin, and charge. As such, QNMs constitute a key probe of black hole properties and play a central role in gravitational wave astronomy~\cite{Berti:2025hly}. The  QNMs may be interpreted  as  massless degrees of freedom  circulating on  null orbits and  gradually leaking away. The orbit's instability timescale is  set by  the corresponding Lyapunov exponent~\cite{Cornish:2003ig,Cardoso:2008bp}
\be
\label{w}
{\rm Im}\,\omega_{\ell}(n)=
\lambda\left(n+\frac{1}{2}\right),
\ee
where $n=0$ defines the fundamental mode and $n\in \mathbb{N}_+$ are the overtones. The Bekenstein-Hod bound in \eq{bekhod} concerns the longest, and therefore physically most relevant, relaxation timescale of the system. In the black hole ringdown, this is set by the least damped QNM. Hence, the bound naturally applies to the fundamental mode, since higher overtones decay faster. 

Therefore, taking $\tau = 1/{\rm Im}\,\omega_{\ell}(0)$,
 \eq{bekhod} becomes identical to \eq{chaos}. 
This indicates that, when applied to the black hole QNMs in the eikonal limit, the Bekenstein-Hod bound is intimately connected to  the  quantum chaos bound on the photon rings.  Moreover, we find that by choosing $T_{\rm ind}$ as the relevant temperature, the Bekenstein-Hod information bound is saturated in the ringdown of black holes in the eikonal limit for the least damped mode, namely for the fundamental mode $n=0$
\be
{\rm Im}\,\omega_{\ell}(0)=\frac{\pi k_B T_{\rm ind}}{\hbar}.
\label{cinqua}
\ee
The Lyapunov exponent sets the timescale over which the black hole relaxes to 
its equilibrium state by emitting QNMs and restoring an effective temperature seen by the local Rindler observer. Although derived in the eikonal limit, \eq{cinqua} is satisfied already at $\ell=4$, as we have checked by comparing our analytical expression for $T_{\rm ind}$ of the equatorial prograde orbit with the numerical results in Ref.~\cite{Berti:2005ys}\footnote{Note that the result in  Ref.~\cite{Hod:2006jw} differs from ours since it refers to the $\ell=2$ mode with the orbit's temperature set to the Hawking temperature, unlike our choice of $T_{\rm ind}$ as the relevant temperature.}. 

The Kerr near-extremal limit deserves a separate comment. In this regime the QNM spectrum contains an infinite tower of highly degenerate overtones, the so-called zero-damped modes
\be
\omega_{\ell}(n)=\frac{1}{2M}\left[\ell-i\epsilon\left(n+\frac{1}{2}\right)\right], ~~~~ \epsilon =\sqrt{1-\frac{a^2}{M^2}}\ll 1,
\ee
and it is not immediately clear which relaxation timescale should be used to test the saturation of the information bound.

Let us consider the physical case of  a massive object orbiting a near-extremal Kerr black hole and quickly  plunging into the
horizon after passing the innermost stable circular orbit. The wavefunctions of the corresponding  outgoing modes are~\cite{Hadar:2014dpa}
\be
h_{n\ell}={\cal A}_\ell\frac{(-i\ell)^n}{n!}
\left(\frac{3\, e^{\epsilon t_0/2M}}{4\,\epsilon}\right)^n\,e^{-i \omega_{\ell}(n)(t-t_0)},
\ee
where ${\cal A}_\ell$ are coefficients that depend on the multipoles. Notice that 
the amplitude of the overtones increases with $n$, a result recently confirmed numerically in Ref.~\cite{DellaRocca:2025zbe}, signaling the fact that the overtones may not be neglected in the near-extremal limit. Nevertheless, the sum over the overtones can be performed to obtain 
\be
\sum_n h_{n\ell}={\cal A}_\ell\,{\rm exp}\left[
 - \frac{\epsilon (t-t_0)}{4M} -i \ell \varphi(t)
\right],~~~~\varphi(t) = \frac{t-t_0}{2M} +\frac{3\, e^{-\frac{\epsilon (t-2t_0)}{2M}}}{4\epsilon},
\ee
from which we deduce that the (inverse of the) relaxation time coincides, after the summation of the overtones, with  the relaxation time of the fundamental mode, 
\be
{\rm Im}\,\omega_{\ell}(n=0)=\frac{\epsilon}{4M}.
\ee
Using Eq.~(\ref{c}), it is easy to see that this expression coincides with  $\pi T_{\rm ind}$ of   the near-extremal Kerr black hole. This result does not come as a surprise because, in the near-extremal Kerr limit, the absolute square of the spectral amplitude  follows the Fermi-Dirac distribution with the temperature  associated with the imaginary part of the fundamental QNM frequency of the fundamental mode~\cite{Oshita:2022pkc}.

Moreover, in the exact extremal case in which the effective temperature vanishes, it has been shown that the real-axis resonances  are unphysical as natural
black hole oscillations: the corresponding frequency is always associated with a scattering mode~\cite{Richartz:2017qep}. This behavior is reminiscent of Nernst’s third law of thermodynamics, which claims that the change in entropy (or lack of information) of a system must vanish in the zero temperature limit.
It seems that once again nature conspires  to  reveal the extreme nature of black holes.

\section{Conclusions}

We have shown that the quantum chaos bound is saturated on the photon rings of black holes. A probe string provides a natural way to uncover the thermal structure of the equatorial near-ring region of Kerr. In the Penrose limit around the equatorial photon ring, a simple static gauge embedding solves the Nambu--Goto equations and induces a genuine two-dimensional Rindler geometry on the string worldsheet. The corresponding Rindler horizon is a worldsheet horizon, and its Unruh temperature defines the induced worldsheet temperature $T_{\rm ind}$.

This offers an interpretation of the saturation of the near-ring chaos bound. The temperature entering the bound is not the Hawking temperature of the Kerr black hole, nor a temperature assigned to a pointlike observer near the photon ring. Rather, it is the Unruh temperature associated with the induced Rindler horizon on the probe
string worldsheet. The surface gravity of this worldsheet Rindler geometry is fixed by the Lyapunov exponent $\lambda$ governing the instability of the equatorial photon ring, realizing  the saturation of the quantum chaos bound in the near-ring region. In this sense, the role played by the black hole horizon in the standard near-horizon case, see \eq{boundsatur}, is replaced here by the photon ring instability.
 
An essential point  of our result is that the four-dimensional instability of the photon ring is encoded in the two-dimensional thermal geometry induced on the probe string. Thus, the photon ring not only is the locus of unstable null geodesics and eikonal ringdown dynamics, but also  gives rise to a thermal Rindler worldsheet whose temperature is directly tied to the associated Lyapunov exponent.
This perspective further highlights the photon ring as a central structure in black hole physics. It is the region where null-orbit instability, resonant wave dynamics, hidden conformal features, and, as shown here, worldsheet thermality and chaos-inspired bounds meet in a common  description. This should not be viewed as a peculiarity of a single black hole solution, but rather as a robust feature of the near-ring black hole geometry.

We have verified the quantum nature of our result by explicitly computing the OTOC functions and showing that the corresponding exponent, $\lambda_L$, equals the geometric Lyapunov exponent $\lambda$. Having OTOCs that grow exponentially with time is one of the criteria to identify quantum chaotic behavior. Krylov complexity (see \cite{Rabinovici:2025otw} for a recent review) is a more refined criterion to test operator growth. It has been claimed in \cite{Parker:2018yvk} that the Krylov complexity  exponent $\alpha$ satisfies $2\alpha\geq \lambda_L$ with equality at saturation. It may be interesting to test this claim for the photon ring system.

Intriguingly, we have shown that the saturation of the near-ring chaos bound is connected  with information emission during the ringdown phase and implies the saturation of the Bekenstein-Hod bound. Indeed, in  the eikonal limit, where QNMs are controlled by photon ring geodesics, the same geometric data that determine the Lyapunov exponent also determine the worldsheet Unruh temperature. This points to a relationship between photon ring instability, thermality, chaos, and the saturation of information bounds such as the Bekenstein bound.

Several directions remain open. It would be important to extend the analysis to non-equatorial photon orbits, where the Penrose-limit profile is more general and the induced worldsheet geometry may be less symmetric.  Finally, it would be interesting to explore how the present picture relates to the recent conjecture that the properties of black holes formed in binary mergers can be inferred from the condition that information spreads from their unstable null orbits at the fastest possible rate~\cite{Giataganas:2026ayt}.

\section*{Acknowledgements}
 We thank Gia Dvali and Julian Sonner for many enlightening discussions. We also thank Emanuele Berti,  Sebastian Cespedes, Shanta de Alwis, Antonio J. Iovino, Andrea Maselli, Harvey Reall, Luciano Rezzolla,  and Jorge Santos for very useful comments. D.G.  acknowledges support from  the National Science and Technology Council (NSTC) of Taiwan with the Young Scholar Columbus Fellowship grant 114-2636-M-110-004.  F.Q.'s research is funded by Tamkeen under the research grant to NYUAD ADHPG-AD457.  A.R.  acknowledges support from the Swiss National Science Foundation (project number CRSII5\_213497).

\appendix
\setcounter{equation}{0}
\setcounter{section}{0}
\setcounter{table}{0}
\makeatletter
\renewcommand{\theequation}{A\arabic{equation}}

\section{Prograde, retrograde and non-equatorial Lyapunov exponents}
\label{App:Aa}
\noindent
In this appendix, we discuss the Lyapunov exponents defined with respect to different time variables, and their extension to non-equatorial spherical photon orbits. Our goal is  to show that, among all possible orbits, the equatorial prograde is the one with the smallest Lyapunov exponent and, therefore, the most relevant for chaotic behavior. 

To make the discussion explicit, we focus on the Kerr black hole, although the construction is expected to generalize to a broad class of stationary axisymmetric black holes of the type considered in this paper. We introduce the dimensionless per-half-orbit instability exponent $\gamma$,  together with the  Lyapunov exponent $\lambda_{\rm p}$, conveniently expressed in Mino time~\cite{Mino:2003yg}, as
\be
\label{eq::Lyapunov_Exponent}
\gamma(r) = \lambda_{\rm p}(r)\, G_\theta(r), ~~~~
G_\theta(r) = \int_{\theta_{-}}^{\theta_+}
\frac{{\rm d}\theta}{\sqrt{\Theta(\theta)}},\qquad \lambda_{\rm p}(r)=
\sqrt{\frac{{\cal R}''(r)}{2}}.
\ee
Here, ${\cal R}(r)$ and $\Theta(\theta)$ denote the radial and polar potentials given by
\begin{align}
    &\mathcal{R}(r)=\left[E (r^2+a^2)- a \ell \right]^2- k \Delta,\qquad 
    \Theta(\theta)= k - \left(\ell \csc\theta - a E \sin \theta\right)^2,\\
     &E= -p_t,\qquad \ell =p_{\phi}=\frac{E\left[M(r^2-a^2)-r\Delta\right]}{a(r-M)}, \\ 
     & k = p_{\theta}^2 + (p_{\phi}\csc \theta + p_t a \sin \theta)^2=\frac{\left[E(r^2+a^2)-a\ell\right]^2}{\Delta},
\end{align}
for spherical null orbits. Moreover, the function $G_\theta$ corresponds to the polar Mino-time period, where Mino time $\tau_{\rm M}$ is defined by  $\dd\tau_{\rm M}= \dd\sigma/\Sigma$, with $\sigma$ being the affine parameter  along the null geodesic.

For non-equatorial spherical photon orbits, the radius remains fixed at $r=r_0$, but the motion in $\theta$ affects the local redshift factor through
\be\la{dtdtau}
\frac{{\rm d}t}{{\rm d}\tau_{\rm M}}=\frac{r^2+a^2}{\Delta}\left[E(r^2+a^2)-a\ell\right]+ a(\ell - aE \sin^2\theta),
\ee
which varies along the trajectory. We can therefore define an instantaneous instability rate $\lambda_t(r_0,\theta)$, which is phase dependent and should be regarded only as a local diagnostic,  rather than the observable asymptotic Lyapunov exponent of the orbit. The only null geodesics for which $\theta$ remains fixed are the equatorial photon rings,  and in that case the local and asymptotic notions coincide.  For generic spherical orbits, the natural orbit-level observable in coordinate time is the $\theta$-averaged exponent, obtained by averaging over the full polar motion and assigning a single constant instability exponent to each orbit. 
\begin{figure*}[t!]
\centering
  \includegraphics[width=0.8\textwidth]{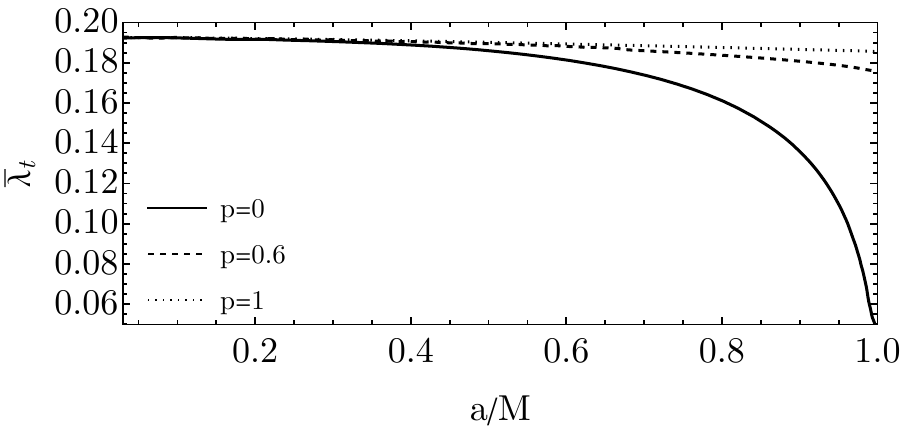}
  \caption{ Orbit-averaged coordinate-time Lyapunov exponent $\overline{\lambda}_t(r)$ across the Kerr photon shell, shown as a function of spin $a/M$ for several fixed shell parameters $p$ with the orbit parametrization as $r=r_- + p(r_+ -r_-)$, so that  $p=0$ corresponds to the prograde equatorial ring, $p=1$ to the retrograde equatorial ring, and $0<p<1$ to non-equatorial spherical photon orbits interpolating between them. The ordering of the curves shows that, for all spins, the prograde branch gives the minimum value of $\overline{\lambda}_t$, while the retrograde branch gives the maximum. As extremality is approached, the prograde exponent is strongly suppressed, whereas the retrograde exponent varies more mildly.}
  \label{fig:2}
\end{figure*}
Consequently, for non-equatorial geodesics we use the phase-dependent local Lyapunov exponent $\lambda_t(r_0,\theta)$. In this section, subscripts indicate the time variable with respect to which the exponent is defined. Thus,
\be
\frac{\dd(\delta r)}{\dd\tau_{\rm M}}=\lambda_{\rm p}(r_0)\delta r,\qquad \frac{\dd(\delta r)}{\dd t}=\lambda_t(r_0,\theta)\delta r~,
\ee
and, by using \eq{dtdtau}, we obtain
\be
\lambda_t(r_0,\theta)=\frac{\lambda_{\rm p}(r_0)}{\dot t(r_0,\theta)}.
\ee
When the equatorial $\lambda_t$ saturates the chaos-bound relation with $T_{\rm ind}$, the Mino-time exponent $\lambda_{\rm p}$ can likewise be associated with an analogous relation involving a rescaled temperature $T_{{\rm p},{\rm eff}}$, obtained from the corresponding change of time coordinate in the two-dimensional Rindler metric.

The orbit-averaged coordinate-time Lyapunov exponent  $\overline{\lambda}_t(r_0)$ is then defined by requiring that, over one polar period, the perturbation grows by the same physical factor  regardless of the time coordinate used to measure it: $\overline{\lambda}_t(r_0) \Delta t= \lambda_{\rm p}(r_0) \Delta \tau_{\rm M}$. It follows that the averaged coordinate-time Lyapunov $\overline{\lambda}_t(r_0)$ is given by
\be
\overline{\lambda}_t(r_0)=\frac{\lambda_{\rm p}(r_0)}{\langle \dot t\rangle_\theta},\qquad \langle \dot t\rangle_\theta
=\frac{ \int_{\theta_-}^{\theta_+} \dot t(r_0,\theta)\,\dd\tau_{\rm M}}
{ \int_{\theta_-}^{\theta_+}\dd\tau_{\rm M}}=
\frac{ \int_{\theta_-}^{\theta_+}\dot t(r_0,\theta)\frac{\dd\theta}{\sqrt{\Theta(\theta)}}}
{ \int_{\theta_-}^{\theta_+}\frac{\dd\theta}{\sqrt{\Theta(\theta)}}}.
\ee
Here we have used
\be
\Delta t=\int_{\theta_-}^{\theta_+} \dd t=\int_{\theta_-}^{\theta_+}\dot t(r_0,\theta)\, \frac{\dd\theta}{\sqrt{\Theta(\theta)}}.
\ee
For the equatorial rings, $\theta=\pi/2$ is fixed, so $\dot t$ is constant. Therefore, $\overline{\lambda}_t(r_\pm)=\lambda_t(r_\pm,\pi/2)=\lambda_\pm $ given by \eq{lambdakerr}. In fact, the equatorial values of $\overline{\lambda}_t$ provide the absolute minimum and maximum of the orbit-averaged exponent for any value of the spin parameter: the prograde branch gives the minimum, while the retrograde branch gives the maximum. This behavior is shown in \fig{fig:2}.

Moreover, one may regard the photon shell as an ensemble of unstable orbits in the interval $r_-\leq r\leq r_+$ with  periodicity $\tau_{\rm M}$. The invariant phase space measure $\tau_{\rm M} \dd r$ then acts as a density of states in the shell, so that orbits with larger Mino-time period contribute more strongly to the expectation values.  The expectation value, or shell-averaged Lyapunov exponent, $\overline{\lambda}_{\rm p}$, can be defined by integrating over all unstable trajectories with respect to the normalized  Mino-time measure~\cite{Giataganas:2026ayt}
\be
\overline{\lambda}_{\rm p}=\frac{\int_{r_-}^{r_+}{\rm d}r \,\gamma(r)}{\int_{r_-}^{r_+}{\rm d}r \,G_\theta(r)}=\frac{\int_{r_-}^{r_+}{\rm d}r \,\lambda_{\rm p}(r)\cdot \tau_{\rm M}(r)}{\int_{r_-}^{r_+}{\rm d}r \,\tau_{\rm M}(r)}.
\ee
The quantity $\overline{\lambda}_{\rm p}$ may thus be interpreted  as the average rate of the total accumulated instability per unit  Mino time. It is expected to be bounded by the retrograde and prograde equatorial Lyapunov exponents, which provide its maximum and minimum values, respectively.

\bibliographystyle{JHEP}
\bibliography{main1}

\providecommand{\href}[2]{#2}\begingroup\raggedright\begin{thebibliography}{10}

\bibitem{LarkinOvchinnikov1969}
A.I.~Larkin and Y.N.~Ovchinnikov, \emph{Quasiclassical method in the theory of superconductivity}, {\emph{Sov. Phys. JETP} {\bfseries 28} (1969) 1200}.

\bibitem{Maldacena:2015waa}
J.~Maldacena, S.H.~Shenker and D.~Stanford, \emph{{A bound on chaos}}, \href{https://doi.org/10.1007/JHEP08(2016)106}{\emph{JHEP} {\bfseries 08} (2016) 106} [\href{https://arxiv.org/abs/1503.01409}{{\ttfamily 1503.01409}}].

\bibitem{Dvali:2013vxa}
G.~Dvali, D.~Flassig, C.~Gomez, A.~Pritzel and N.~Wintergerst, \emph{{Scrambling in the Black Hole Portrait}}, \href{https://doi.org/10.1103/PhysRevD.88.124041}{\emph{Phys. Rev. D} {\bfseries 88} (2013) 124041} [\href{https://arxiv.org/abs/1307.3458}{{\ttfamily 1307.3458}}].

\bibitem{Shenker:2013pqa}
S.H.~Shenker and D.~Stanford, \emph{{Black holes and the butterfly effect}}, \href{https://doi.org/10.1007/JHEP03(2014)067}{\emph{JHEP} {\bfseries 03} (2014) 067} [\href{https://arxiv.org/abs/1306.0622}{{\ttfamily 1306.0622}}].

\bibitem{Polchinski:2015cea}
J.~Polchinski, \emph{{Chaos in the black hole S-matrix}},  \href{https://arxiv.org/abs/1505.08108}{{\ttfamily 1505.08108}}.

\bibitem{Hayden:2007cs}
P.~Hayden and J.~Preskill, \emph{{Black holes as mirrors: Quantum information in random subsystems}}, \href{https://doi.org/10.1088/1126-6708/2007/09/120}{\emph{JHEP} {\bfseries 09} (2007) 120} [\href{https://arxiv.org/abs/0708.4025}{{\ttfamily 0708.4025}}].

\bibitem{Sekino:2008he}
Y.~Sekino and L.~Susskind, \emph{{Fast Scramblers}}, \href{https://doi.org/10.1088/1126-6708/2008/10/065}{\emph{JHEP} {\bfseries 10} (2008) 065} [\href{https://arxiv.org/abs/0808.2096}{{\ttfamily 0808.2096}}].

\bibitem{Hadar:2022xag}
S.~Hadar, D.~Kapec, A.~Lupsasca and A.~Strominger, \emph{{Holography of the photon ring}}, \href{https://doi.org/10.1088/1361-6382/ac8d43}{\emph{Class. Quant. Grav.} {\bfseries 39} (2022) 215001} [\href{https://arxiv.org/abs/2205.05064}{{\ttfamily 2205.05064}}].

\bibitem{Kapec:2022dvc}
D.~Kapec, A.~Lupsasca and A.~Strominger, \emph{{Photon rings around warped black holes}}, \href{https://doi.org/10.1088/1361-6382/acc164}{\emph{Class. Quant. Grav.} {\bfseries 40} (2023) 095006} [\href{https://arxiv.org/abs/2211.01674}{{\ttfamily 2211.01674}}].

\bibitem{Giataganas:2024hil}
D.~Giataganas, A.~Kehagias and A.~Riotto, \emph{{Quasinormal modes and universality of the Penrose limit of black hole photon rings}}, \href{https://doi.org/10.1007/JHEP09(2024)168}{\emph{JHEP} {\bfseries 09} (2024) 168} [\href{https://arxiv.org/abs/2403.10605}{{\ttfamily 2403.10605}}].

\bibitem{Giataganas:2026ayt}
D.~Giataganas, G.F.~Giudice, A.~Ianniccari, A.J.~Iovino, A.~Kehagias, F.~Quevedo et~al., \emph{{Black Hole Mergers as the Fastest Photon Ring Scramblers}},  \href{https://arxiv.org/abs/2603.08643}{{\ttfamily 2603.08643}}.

\bibitem{EventHorizonTelescope:2019dse}
{\scshape Event Horizon Telescope} collaboration, \emph{{First M87 Event Horizon Telescope Results. I. The Shadow of the Supermassive Black Hole}}, \href{https://doi.org/10.3847/2041-8213/ab0ec7}{\emph{Astrophys. J. Lett.} {\bfseries 875} (2019) L1} [\href{https://arxiv.org/abs/1906.11238}{{\ttfamily 1906.11238}}].

\bibitem{Chael:2021rjo}
A.~Chael, M.D.~Johnson and A.~Lupsasca, \emph{{Observing the Inner Shadow of a Black Hole: A Direct View of the Event Horizon}}, \href{https://doi.org/10.3847/1538-4357/ac09ee}{\emph{Astrophys. J.} {\bfseries 918} (2021) 6} [\href{https://arxiv.org/abs/2106.00683}{{\ttfamily 2106.00683}}].

\bibitem{Johnson:2019ljv}
M.D.~Johnson et~al., \emph{{Universal interferometric signatures of a black hole{\textquoteright}s photon ring}}, \href{https://doi.org/10.1126/sciadv.aaz1310}{\emph{Sci. Adv.} {\bfseries 6} (2020) eaaz1310} [\href{https://arxiv.org/abs/1907.04329}{{\ttfamily 1907.04329}}].

\bibitem{blau}
M.~Blau, \emph{{Plane Waves and Penrose Limits}},  \href{https://arxiv.org/abs/http://www.blau.itp.unibe.ch/Lecturenotes.html}{{\ttfamily http://www.blau.itp.unibe.ch/Lecturenotes.html}}.

\bibitem{Fransen:2023eqj}
K.~Fransen, \emph{{Quasinormal modes from Penrose limits}}, \href{https://doi.org/10.1088/1361-6382/acf26d}{\emph{Class. Quant. Grav.} {\bfseries 40} (2023) 205004} [\href{https://arxiv.org/abs/2301.06999}{{\ttfamily 2301.06999}}].

\bibitem{Cardoso:2008bp}
V.~Cardoso, A.S.~Miranda, E.~Berti, H.~Witek and V.T.~Zanchin, \emph{{Geodesic stability, Lyapunov exponents and quasinormal modes}}, \href{https://doi.org/10.1103/PhysRevD.79.064016}{\emph{Phys. Rev. D} {\bfseries 79} (2009) 064016} [\href{https://arxiv.org/abs/0812.1806}{{\ttfamily 0812.1806}}].

\bibitem{Deich:2023oox}
A.~Deich, N.~Yunes and C.~Gammie, \emph{{Lyapunov exponents to test general relativity}}, \href{https://doi.org/10.1103/PhysRevD.110.044033}{\emph{Phys. Rev. D} {\bfseries 110} (2024) 044033} [\href{https://arxiv.org/abs/2308.07232}{{\ttfamily 2308.07232}}].

\bibitem{Horowitz:1990sr}
G.T.~Horowitz and A.R.~Steif, \emph{{Strings in Strong Gravitational Fields}}, \href{https://doi.org/10.1103/PhysRevD.42.1950}{\emph{Phys. Rev. D} {\bfseries 42} (1990) 1950}.

\bibitem{Dodelson:2015toa}
M.~Dodelson and E.~Silverstein, \emph{{String-theoretic breakdown of effective field theory near black hole horizons}}, \href{https://doi.org/10.1103/PhysRevD.96.066010}{\emph{Phys. Rev. D} {\bfseries 96} (2017) 066010} [\href{https://arxiv.org/abs/1504.05536}{{\ttfamily 1504.05536}}].

\bibitem{Nakamura:2013yqa}
S.~Nakamura and H.~Ooguri, \emph{{Out of Equilibrium Temperature from Holography}}, \href{https://doi.org/10.1103/PhysRevD.88.126003}{\emph{Phys. Rev. D} {\bfseries 88} (2013) 126003} [\href{https://arxiv.org/abs/1309.4089}{{\ttfamily 1309.4089}}].

\bibitem{Sonner:2012if}
J.~Sonner and A.G.~Green, \emph{{Hawking Radiation and Non-equilibrium Quantum Critical Current Noise}}, \href{https://doi.org/10.1103/PhysRevLett.109.091601}{\emph{Phys. Rev. Lett.} {\bfseries 109} (2012) 091601} [\href{https://arxiv.org/abs/1203.4908}{{\ttfamily 1203.4908}}].

\bibitem{Giataganas:2013hwa}
D.~Giataganas and H.~Soltanpanahi, \emph{{Universal Properties of the Langevin Diffusion Coefficients}}, \href{https://doi.org/10.1103/PhysRevD.89.026011}{\emph{Phys. Rev. D} {\bfseries 89} (2014) 026011} [\href{https://arxiv.org/abs/1310.6725}{{\ttfamily 1310.6725}}].

\bibitem{Raffaelli:2021gzh}
B.~Raffaelli, \emph{{Hidden conformal symmetry on the black hole photon sphere}}, \href{https://doi.org/10.1007/JHEP03(2022)125}{\emph{JHEP} {\bfseries 03} (2022) 125} [\href{https://arxiv.org/abs/2112.12543}{{\ttfamily 2112.12543}}].

\bibitem{Bisognano:1975ih}
J.J.~Bisognano and E.H.~Wichmann, \emph{{On the Duality Condition for a Hermitian Scalar Field}}, \href{https://doi.org/10.1063/1.522605}{\emph{J. Math. Phys.} {\bfseries 16} (1975) 985}.

\bibitem{Bisognano:1976za}
J.J.~Bisognano and E.H.~Wichmann, \emph{{On the Duality Condition for Quantum Fields}}, \href{https://doi.org/10.1063/1.522898}{\emph{J. Math. Phys.} {\bfseries 17} (1976) 303}.

\bibitem{Crispino:2007eb}
L.C.B.~Crispino, A.~Higuchi and G.E.A.~Matsas, \emph{{The Unruh effect and its applications}}, \href{https://doi.org/10.1103/RevModPhys.80.787}{\emph{Rev. Mod. Phys.} {\bfseries 80} (2008) 787} [\href{https://arxiv.org/abs/0710.5373}{{\ttfamily 0710.5373}}].

\bibitem{Perrone:2025zhy}
D.~Perrone, A.~Kehagias and A.~Riotto, \emph{{Nonlinearities in Kerr black hole ringdown from the Penrose limit}}, \href{https://doi.org/10.1088/1475-7516/2025/10/024}{\emph{JCAP} {\bfseries 10} (2025) 024} [\href{https://arxiv.org/abs/2507.01919}{{\ttfamily 2507.01919}}].

\bibitem{Hashimoto:2017oit}
K.~Hashimoto, K.~Murata and R.~Yoshii, \emph{{Out-of-time-order correlators in quantum mechanics}}, \href{https://doi.org/10.1007/JHEP10(2017)138}{\emph{JHEP} {\bfseries 10} (2017) 138} [\href{https://arxiv.org/abs/1703.09435}{{\ttfamily 1703.09435}}].

\bibitem{Roberts:2014isa}
D.A.~Roberts, D.~Stanford and L.~Susskind, \emph{{Localized shocks}}, \href{https://doi.org/10.1007/JHEP03(2015)051}{\emph{JHEP} {\bfseries 03} (2015) 051} [\href{https://arxiv.org/abs/1409.8180}{{\ttfamily 1409.8180}}].

\bibitem{Takahashi:1996zn}
Y.~Takahashi and H.~Umezawa, \emph{{Thermo field dynamics}}, \href{https://doi.org/10.1142/S0217979296000817}{\emph{Int. J. Mod. Phys. B} {\bfseries 10} (1996) 1755}.

\bibitem{Israel:1976ur}
W.~Israel, \emph{{Thermo field dynamics of black holes}}, \href{https://doi.org/10.1016/0375-9601(76)90178-X}{\emph{Phys. Lett. A} {\bfseries 57} (1976) 107}.

\bibitem{Maldacena:2001kr}
J.M.~Maldacena, \emph{{Eternal black holes in anti-de Sitter}}, \href{https://doi.org/10.1088/1126-6708/2003/04/021}{\emph{JHEP} {\bfseries 04} (2003) 021} [\href{https://arxiv.org/abs/hep-th/0106112}{{\ttfamily hep-th/0106112}}].

\bibitem{Dray:1984ha}
T.~Dray and G.~'t~Hooft, \emph{{The Gravitational Shock Wave of a Massless Particle}}, \href{https://doi.org/10.1016/0550-3213(85)90525-5}{\emph{Nucl. Phys. B} {\bfseries 253} (1985) 173}.

\bibitem{Camanho:2014apa}
X.O.~Camanho, J.D.~Edelstein, J.~Maldacena and A.~Zhiboedov, \emph{{Causality Constraints on Corrections to the Graviton Three-Point Coupling}}, \href{https://doi.org/10.1007/JHEP02(2016)020}{\emph{JHEP} {\bfseries 02} (2016) 020} [\href{https://arxiv.org/abs/1407.5597}{{\ttfamily 1407.5597}}].

\bibitem{Araneda:2022lgu}
B.~Araneda, \emph{{Parallel spinors, pp-waves, and gravitational perturbations}}, \href{https://doi.org/10.1088/1361-6382/acaa83}{\emph{Class. Quant. Grav.} {\bfseries 40} (2023) 025006} [\href{https://arxiv.org/abs/2204.13673}{{\ttfamily 2204.13673}}].

\bibitem{Kehagias:2024sgh}
A.~Kehagias and A.~Riotto, \emph{{Nonlinear effects in black hole ringdown made simple: Quasinormal modes as adiabatic modes}}, \href{https://doi.org/10.1103/PhysRevD.111.L041506}{\emph{Phys. Rev. D} {\bfseries 111} (2025) L041506} [\href{https://arxiv.org/abs/2411.07980}{{\ttfamily 2411.07980}}].

\bibitem{Bekenstein:1981zz}
J.D.~Bekenstein, \emph{{Energy Cost of Information Transfer}}, \href{https://doi.org/10.1103/PhysRevLett.46.623}{\emph{Phys. Rev. Lett.} {\bfseries 46} (1981) 623}.

\bibitem{Hod:2006jw}
S.~Hod, \emph{{Universal Bound on Dynamical Relaxation Times and Black-Hole Quasinormal Ringing}}, \href{https://doi.org/10.1103/PhysRevD.75.064013}{\emph{Phys. Rev. D} {\bfseries 75} (2007) 064013} [\href{https://arxiv.org/abs/gr-qc/0611004}{{\ttfamily gr-qc/0611004}}].

\bibitem{Berti:2025hly}
J.~Abedi et~al., \emph{{Black hole spectroscopy: from theory to experiment}},  \href{https://arxiv.org/abs/2505.23895}{{\ttfamily 2505.23895}}.

\bibitem{Cornish:2003ig}
N.J.~Cornish and J.J.~Levin, \emph{{Lyapunov timescales and black hole binaries}}, \href{https://doi.org/10.1088/0264-9381/20/9/304}{\emph{Class. Quant. Grav.} {\bfseries 20} (2003) 1649} [\href{https://arxiv.org/abs/gr-qc/0304056}{{\ttfamily gr-qc/0304056}}].

\bibitem{Berti:2005ys}
E.~Berti, V.~Cardoso and C.M.~Will, \emph{{On gravitational-wave spectroscopy of massive black holes with the space interferometer LISA}}, \href{https://doi.org/10.1103/PhysRevD.73.064030}{\emph{Phys. Rev. D} {\bfseries 73} (2006) 064030} [\href{https://arxiv.org/abs/gr-qc/0512160}{{\ttfamily gr-qc/0512160}}].

\bibitem{Hadar:2014dpa}
S.~Hadar, A.P.~Porfyriadis and A.~Strominger, \emph{{Gravity Waves from Extreme-Mass-Ratio Plunges into Kerr Black Holes}}, \href{https://doi.org/10.1103/PhysRevD.90.064045}{\emph{Phys. Rev. D} {\bfseries 90} (2014) 064045} [\href{https://arxiv.org/abs/1403.2797}{{\ttfamily 1403.2797}}].

\bibitem{DellaRocca:2025zbe}
M.~Della~Rocca, L.~Pezzella, E.~Berti, L.~Gualtieri and A.~Maselli, \emph{{Quasinormal ringing of Kerr black holes. III. Excitation coefficients for equatorial inspirals from the innermost stable circular orbit}},  \href{https://arxiv.org/abs/2512.07959}{{\ttfamily 2512.07959}}.

\bibitem{Oshita:2022pkc}
N.~Oshita, \emph{{Thermal ringdown of a Kerr black hole: overtone excitation, Fermi-Dirac statistics and greybody factor}}, \href{https://doi.org/10.1088/1475-7516/2023/04/013}{\emph{JCAP} {\bfseries 04} (2023) 013} [\href{https://arxiv.org/abs/2208.02923}{{\ttfamily 2208.02923}}].

\bibitem{Richartz:2017qep}
M.~Richartz, C.A.R.~Herdeiro and E.~Berti, \emph{{Synchronous frequencies of extremal Kerr black holes: resonances, scattering and stability}}, \href{https://doi.org/10.1103/PhysRevD.96.044034}{\emph{Phys. Rev. D} {\bfseries 96} (2017) 044034} [\href{https://arxiv.org/abs/1706.01112}{{\ttfamily 1706.01112}}].

\bibitem{Rabinovici:2025otw}
E.~Rabinovici, A.~S{\'a}nchez-Garrido, R.~Shir and J.~Sonner, \emph{{Krylov Complexity}},  \href{https://arxiv.org/abs/2507.06286}{{\ttfamily 2507.06286}}.

\bibitem{Parker:2018yvk}
D.E.~Parker, X.~Cao, A.~Avdoshkin, T.~Scaffidi and E.~Altman, \emph{{A Universal Operator Growth Hypothesis}}, \href{https://doi.org/10.1103/PhysRevX.9.041017}{\emph{Phys. Rev. X} {\bfseries 9} (2019) 041017} [\href{https://arxiv.org/abs/1812.08657}{{\ttfamily 1812.08657}}].

\bibitem{Mino:2003yg}
Y.~Mino, \emph{{Perturbative approach to an orbital evolution around a supermassive black hole}}, \href{https://doi.org/10.1103/PhysRevD.67.084027}{\emph{Phys. Rev. D} {\bfseries 67} (2003) 084027} [\href{https://arxiv.org/abs/gr-qc/0302075}{{\ttfamily gr-qc/0302075}}].

\end{thebibliography}\endgroup

\end{document}